\documentclass{aa} 
\newcommand{\appropto}{\mathrel{\vcenter{
  \offinterlineskip\halign{\hfil$##$\cr
    \propto\cr\noalign{\kern2pt}\sim\cr\noalign{\kern-2pt}}}}}
\usepackage{graphicx}
\usepackage[varg]{txfonts}
\usepackage{mathrsfs}
\usepackage[flushleft]{threeparttable}
\usepackage[colorlinks=true,allcolors=blue]{hyperref}
\begin{document}

   \title{$\beta$ Pictoris b through the eyes of the upgraded CRIRES+}
   \subtitle{Atmospheric composition, spin rotation, and radial velocity
   \thanks{Based on observations collected at the European Southern Observatory under ESO programme 108.22HG.001}}

    \author{R. Landman\inst{1}, T. Stolker\inst{1}, I.A.G. Snellen\inst{1}, J. Costes \inst{2}, S. de Regt\inst{1}, Y. Zhang\inst{1}, \\S. Gandhi\inst{1,3}, P. Molliere \inst{4}, A. Kesseli\inst{5}, A. Vigan\inst{2}, A. S\'{a}nchez-L\'{o}pez\inst{6}}
    \authorrunning{R. Landman et al.}

   \institute{Leiden Observatory, Leiden University, P.O. Box 9513, 2300 RA Leiden, The Netherlands \\
   \email{rlandman@strw.leidenuniv.nl}
   \and
   Aix Marseille Univ., CNRS, CNES, LAM, Marseille, France
   \and
   Department of Physics, University of Warwick, Coventry CV4 7AL, UK.   3. Centre for Exoplanets and Habitability, University of Warwick, Gibbet Hill Road, Coventry CV4 7AL, UK
   \and
   Max-Planck Institut für Astronomie, Königstuhl17, 69117 Heidelberg, Germany
   \and
   IPAC, Mail Code 100-22, Caltech, 1200 E. California Boulevard, Pasadena, CA 91125, USA
   \and
   Instituto de Astrof{\'i}sica de Andaluc{\'i}a (IAA-CSIC), Glorieta de la Astronom{\'i}a s/n, 18008 Granada, Spain
   }

   \date{Received XXX; accepted YYY}

% \abstract{}{}{}{}{} 
% 5 {} token are mandatory
 
  \abstract
    {High-resolution spectrographs fed by adaptive optics (AO) provide a unique opportunity to characterize directly imaged exoplanets. Observations with such instruments allow us to probe the atmospheric composition, spin rotation, and radial velocity of the planet, thereby helping to reveal information on its formation and migration history. The recent upgrade of the Cryogenic High-Resolution Infrared Echelle Spectrograph (CRIRES+) at the VLT makes it a highly suitable instrument for characterizing directly imaged exoplanets.}
   {In this work, we report on observations of $\beta$ Pictoris b with CRIRES+ and use them to constrain the planets atmospheric properties and update the estimation of its spin rotation.}
  % methods heading (mandatory)  m
   {The data were reduced using the open-source \textit{pycrires} package. We subsequently forward-modeled the stellar, planetary, and systematic contribution to the data to detect molecules in the planet's atmosphere. We also used atmospheric retrievals to provide new constraints on its atmosphere.}
   {We confidently detected water and carbon monoxide in the atmosphere of $\beta$ Pictoris b and retrieved a slightly sub-solar carbon-to-oxygen ratio, which is in agreement with previous results. The interpretation is hampered by our limited knowledge of the C/O ratio of the host star. We also obtained a much improved constraint on its spin rotation of $19.9 \pm 1.0$ km/s, which gives a rotation period of $8.7 \pm 0.8$ hours, assuming no obliquity. We find that there is a degeneracy between the metallicity and clouds, but this has minimal impact on the retrieved C/O, $v\sin{i}$, and radial velocity. Our results show that CRIRES+ is performing well and stands as a highly useful instrument for characterizing directly imaged planets.}
   {}


    \maketitle
 
%
%-------------------------------------------------------------------

\section{Introduction}

The growing group of directly imaged super-Jupiters at wide separations continue to pose a challenge to planet formation models. It is unclear whether these planets form bottom-up via core accretion or represent the tail end of star formation and form top-down via gravitational instability. Occurrence rates from large surveys have started to provide constraints on the population as a whole \citep{2019AJ_nielsen_gpies, 2021A&A_vigan_shine}, offering evidence of multiple formation pathways. However, understanding the formation and evolution history of individual objects remains a challenge. Furthermore, there is still a great deal of uncertainty around the chemical and physical processes in the atmospheres of these planets, namely, the possible presence and composition of (patchy) clouds or disequilibrium chemistry \cite[e.g.,][]{Rajan2017AJ_51eri,2020A&A_molliere_hr8799}. Since we are able to spatially resolve them from their host star, these young, self-luminous planets at wide separations are well suited for atmospheric characterizations, while establishing links between their atmospheric properties and their formation and migration histories. 

Multiple tracers of the planet's formation and migration history have been proposed, which are accessible with present-day observations. First, the elemental and isotopic abundance ratios in the planets atmosphere, such as the carbon-to-oxygen (C/O) ratio, have been advocated as useful tracers of where the planet has accreted its gas in the protoplanetary disk  \citep{Oberg2011ApJ_icelines, Madhusudhan2017MNRAS_C_O_formation, 2021Natur_zhang_isotope}. After measuring such abundance ratios, we can attempt to invert this problem and constrain the planet's formation and migration history \citep{Gravity2020A&A_betapicb,Molliere2022arXiv_formation_composition}. The C/O ratio has been measured for a growing number of directly imaged planets and brown dwarf companions \cite[e.g.][see \citet{Hoch2023AJ_co_ratio} for an overview]{2013Sci_konopacky_hr8799_osiris, 2020A&A_molliere_hr8799,Gravity2020A&A_betapicb,2020AJ_wilcomb_kappa_and_b_osiris, Wang_2021AJ_hr8799_kpic, Wang2021AJ_PDS70_gravity, 2021A&A_petrus_hip65426b, 2022AJHoch_VHS1256_OSIRIS, Xuan2022ApJ_browndwarf_kpic, PalmaBifani2023A&A_ABPic_sinfoni, BrownSevilla2023A&A_51Eri,Wang2023AJ_Hr8799c_hrs_lrs} as well as hot Jupiters \citep[e.g.,][]{Madhusudhan2011Natu_C_O_wasp12b,Line2021Natur_C_O_wasp77Ab, Changeat2022ApJS_25hotjupiters}. While low-resolution spectroscopy generally results in large uncertainties due to degeneracies with, for instance, clouds and surface gravity, medium-to-high-resolution spectroscopy has resulted in accurate and robust estimates of the C/O ratio for a handful of directly imaged planets \cite[e.g.,][]{Ruffio_2021AJ_HR8799_deep, Wang_2021AJ_hr8799_kpic, 2021Natur_zhang_isotope}.
Second, the planets' spin rotation data contain information on the evolution of their angular momentum. This rotation rate may deviate depending on the formation channel or through interactions with the circumplanetary disk. \citet{Bryan2018NatAs_spin_planetary_mass} found no distinction in the distribution of spin measurements for planetary mass companions and isolated brown dwarfs, indicating that their spin evolution is regulated in the same way. On the other hand, \citet{Wang_2021AJ_hr8799_kpic} found a tentative trend in increasing rotation rate with decreasing companion mass, which could indicate that magnetic braking is less efficient for lower mass companions. While the rotation rates of a handful of planetary mass companions have been measured \citep{2014Natur_snellen_beta_pic_crires,2016A&A_schwarz_spin_gq_lup, Bryan2018NatAs_spin_planetary_mass,Xuan2020AJ_DHTaub,Wang_2021AJ_hr8799_kpic}, a larger sample and more accurate measurements are needed to confirm this trend.
Finally, the orbit of the planet can reveal its dynamical history and whether it has, for example, experienced a scattering event. High-resolution spectroscopy allows us to measure the radial velocity of the planet, which can be used to break degeneracies in orbital fits \cite[e.g.,][]{2016A&A_schwarz_spin_gq_lup, Ruffio_2021AJ_HR8799_deep}.

High-resolution spectrographs equipped with adaptive optics (AO) provide the unique capability of measuring robust elemental abundance ratios, spin rotation, and radial velocity for directly imaged planets. One of the best targets for this is the emblematic $\beta$ Pictoris b \citep{2010Sci_lagrange_beta_pic} due to its close proximity. Its spin rotation was estimated using the original CRIRES by \citet{2014Natur_snellen_beta_pic_crires}, showing a fast rotation rate of $v\sin i = 25 \pm 3$ km/s. Many works have characterized its atmosphere \citep[e.g.,][]{Bonnefoy2013A&A_betaPicb,Currie2013ApJ_betaPicb, Morzinski2015ApJ_betaPicb_MagAO, Baudino2015A&A_betapicb,2018A&A_hoeijmakers_mm, Stolker2020A&A_miracles}, with most studies finding effective temperatures of 1650-1800 K, low surface gravities, and thick clouds. The most detailed study to date was performed by \citet{Gravity2020A&A_betapicb}, who report a sub-solar C/O ratio of $0.43 \pm 0.05$ from atmospheric retrievals on VLTI/GRAVITY K-band data in combination with low-spectral resolution YJH data from GPI \citep{Chilcote2017AJ_beta_pic_b_gpi}. Assuming that the host star $\beta$ Pictoris has a solar C/O ratio, the authors concluded that the planet has most likely formed via core accretion. 

In this work, we report on observations with the newly refurbished VLT/CRIRES+ instrument to probe the atmosphere of $\beta$ Pictoris b and provide an updated measurement of its spin rotation and radial velocity. Section \ref{sec:observations} describes the observations and data reduction pipeline. Section \ref{sec:methods} details the data modeling of the observations and Section \ref{sec:atmosphere_models} describes the atmospheric models of the planet. Section \ref{sec:results} reports the detection of molecules and results from atmospheric retrievals, which are discussed in Section \ref{sec:discussion}. Finally, our conclusions are summarized in Section \ref{sec:conclusions}. 

\section{Observations and data reduction}\label{sec:observations}
\subsection{Observations}
We observed the $\beta$ Pictoris system for 40 minutes of integration time on both 11 November 2021 and 13 November 2021 with the upgraded Cryogenic High-Resolution Infrared Echelle Spectrograph (CRIRES+; \citet{Dorn2014Msngr_crires+, Dorn2023A&A_crires+}) as part of program 108.22HG.001. CRIRES+ is located at the  Nasmyth focus B of UT 3 at the VLT and has been upgraded with a cross-disperser and three Hawaii 2RG detectors, increasing the simultaneous wavelength coverage by up to a factor ten. We used wavelength setting K2166 and a slit width of 0.2", resulting in a spectral resolving power of $R\sim 100,000$. Throughout this work, we do not include the first two of the seven spectral orders, as these are heavily dominated by tellurics. This results in an effective wavelength coverage of 2.06-2.47 $\mu m$, but with gaps in between due to the different spectral orders. The slit was placed in such a way that it encompassed both the host star and $\beta$ Pictoris b. The presence of the star on the slit significantly limited the maximum exposure time in order to avoid saturation. At the time of observation the planet was at a separation of 0.48" and a position angle of 31.25 degrees \citep{Wang2021ascl.soft_whereistheplanet}. The observations were taken using ABBA nodding with DIT=15s and NDIT=8, resulting in 10 frames in both nodding positions for both nights. The observations were taken with adaptive optics from MACAO and the conditions were decent with seeing ranging from 0.55" to 1" during the observations. 

\subsection{Data reduction with \textit{pycrires}}
\begin{figure}[htbp]
    \centering
    \includegraphics[width=\linewidth]{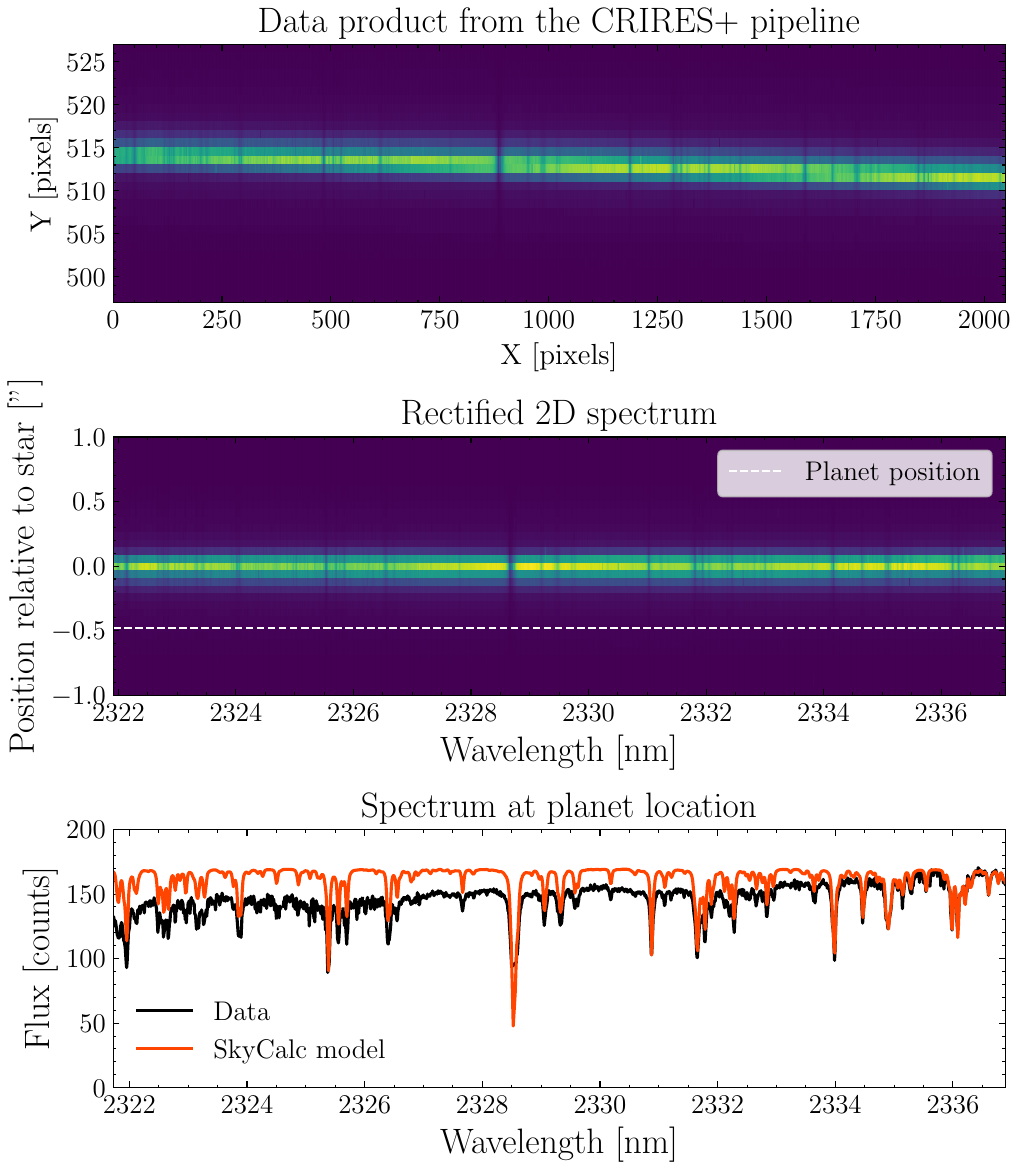}
    \caption{Visualization of the data extraction pipeline. Top row shows the product of the \textit{obs\_nodding} recipe from the ESO CRIRES+ data reduction pipeline. The middle row shows the rectified 2D spectrum corrected for the slit tilt and curvature using \textit{pycrires} and the bottom row shows the spectrum at the location of the planet, plotted together with a telluric model generated with SkyCalc.}
    \label{fig:data_reduction}
\end{figure}

The raw data was reduced using the \textit{pycrires}\footnote{\url{https://github.com/tomasstolker/pycrires}} package for python \citep{Stolker2023ascl.sof_pycrires}. This package contains python wrappers for the CRIRES+ EsoRex pipeline and custom functions specifically developed for direct observations of substellar companions. We used the EsoRex pipeline version 1.2 to perform the basic image processing, including dark and flat field correction and an initial wavelength solution using the uranium-neon lamp and Fabry-Perot etalon calibration files \citep{Seemann2014SPIE_crires+_wavecal}. We then used the intermediate data products from the \textit{obs\_nodding} function, which was run on each nodding pair separately, as input to our custom functions. This data product is shown in the top panel of Fig. \ref{fig:data_reduction}. We subsequently correct for the curvature of the slit on the detector and the slit tilt using interpolation and the information from the trace-wave table produced by the EsoRex pipeline to obtain 2D rectified spectra, as shown in the central panel of Fig. \ref{fig:data_reduction}. We found that there were slight vertical shifts of about a pixel between different spectral orders in these 2D rectified spectra, likely due to slight errors in the trace determination. To remove this, we fit a Gaussian to the slit illumination, which was calculated by summing the 2D rectified spectra along the spectral dimension. This was done for each spectral order and exposure and we subsequently aligned all the orders using interpolation.

We found the wavelength solution from the EsoRex pipeline to be not sufficiently accurate, so we performed an additional correction on the wavelength solution in a similar way as \citet{2022AJ_Holmberg_crires+}. We fitted for a second order polynomial correction to the current best wavelength solution, which is initially obtained from the CRIRES+ pipeline. This was done by maximizing the cross-correlation between the extracted and continuum removed spectrum of $\beta$ Pic and a template of the telluric absorption generated with SkyCalc \citep{2012A&A_Noll_skycalc} on a grid of polynomial coefficients. We repeated this process three times and gradually increased the accuracy of the grid around the current best wavelength solution. We found that this resulted in accurate wavelength solutions, even in orders where there are minimal tellurics, which we checked by visual comparison of the resulting spectrum and a telluric model, and through inspection of the resulting cross-correlation functions. An example of the observed spectrum at the planet's location overlayed with the telluric model is shown in the bottom panel of Fig. \ref{fig:data_reduction}.

\section{Analysis framework}\label{sec:methods}
\begin{figure}[htbp]
    \centering
    \includegraphics[width=\linewidth]{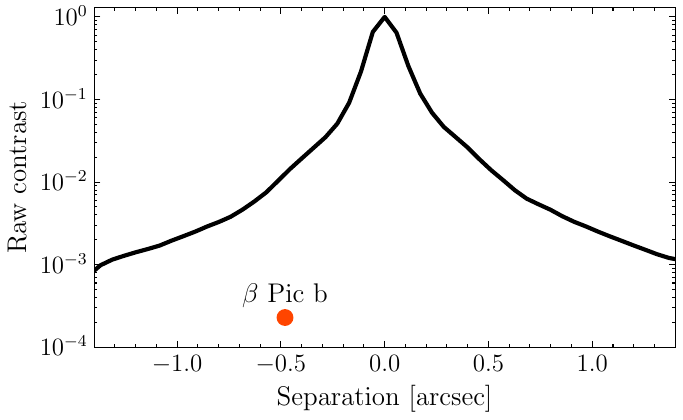}
    \caption{Raw white light contrast curve for our CRIRES+ observations. The location of $\beta$ Pictoris b is indicated in this diagram according to the K-band contrast from \citet{Chilcote2017AJ_beta_pic_b_gpi}.}
    \label{fig:contrast_curve}
\end{figure}

While the adaptive optics performance was good, the signal at the location of the planet is dominated by the stellar contribution. From the reduced data, we measure the stellar flux to be a factor 100 lower at the location of the planet position compared to the peak stellar flux, as shown in the raw broad-band contrast curve in Fig. \ref{fig:contrast_curve}. Given the star-planet contrast of $\Delta K = 9.1$ \citep{Chilcote2017AJ_beta_pic_b_gpi}, this means that the planet only contributes about $\sim$2\% of the flux at its position. It is therefore not straightforward to extract the planet spectrum and derive its atmospheric parameters. In this section, we explain our data modeling, which closely follows the forward modeling approach developed in \citet{2019AJ_ruffio_radial_velocity, Ruffio_2021AJ_HR8799_deep} for OSIRIS data of the HR8799 planets. Even though the framework from \citet{2019AJ_ruffio_radial_velocity} was developed for medium spectral resolution data from an integral field spectrograph, it is almost directly applicable to the high-resolution long-slit data from CRIRES+.

\subsection{Data modeling}
The reduced 1D spectrum, $\textbf{d}$, at a specific position along the slit can be written as a vector with (as entries) the flux at each wavelength $\lambda$. This signal is a combination of the stellar signal, $\textbf{d}_s$; planet signal, $\textbf{d}_p$; systematics, $\textbf{d}_{\textrm{sys}}$; and noise, $\eta$. It can be written as:
\begin{equation}\label{eq:base_model}
    \textbf{d} = \textbf{d}_s + \textbf{d}_p + \textbf{d}_{\textrm{sys}} + \eta.
\end{equation}
To derive parameters of the planet's atmosphere, we forward-modeled these contributions. We expressed the observed data as a linear combination of stellar, planetary, and systematic components, closely following \citet{2019AJ_ruffio_radial_velocity}:
\begin{equation}\label{eq:linear_model}
    \textbf{d} = M_{\psi}\textbf{c} + \eta,
\end{equation}
where $M_{\psi}$ is the linear model matrix with nonlinear parameters, $\psi$, and $\textbf{c}$ is a vector with entries corresponding to the amplitude of each linear component, and $\eta$ is the noise. The different components that make up $M_\psi$ are individually explained in the next subsections.

\subsubsection{Stellar contribution}\label{sec:stellar_model}
We estimated the telluric imprinted stellar master spectrum, $\mathbf{f}_s$, using the on-axis observation of the star. To reduce the influence of noisy datapoints, we estimated this stellar master spectrum by taking the sum over the five rows centered on the host star, which are free from planet signal. However, because speckles move outwards for increasing wavelengths, this stellar contribution is modulated by a low-order function $\alpha(\lambda)$ off-axis:
\begin{equation}
    \textbf{d}_s = c_s \alpha(\lambda) \textbf{f}_s,
\end{equation}
where $c_s$ is a unitless linear scaling parameter, part of $\textbf{c}$ in Eq. \ref{eq:linear_model}, which we fit for. One can choose how to parameterize the modulation, $\alpha(\lambda)$. In principle, we can retain the continuum from the planet by jointly fitting the planet signal and this modulation \citep{Ruffio2023AJ_exomoons}. However, since we are completely dominated by the stellar contribution, which also has almost no spectral lines, this is not possible in our case. Instead, we estimated $\alpha$ directly from the data using:

\begin{equation}\label{eq:alpha_est}
    \hat{\alpha}(\lambda) = \frac{\mathcal{L}[\textbf{d}_s]} {\mathcal{L}[\textbf{f}_s]},
\end{equation}
where $\mathcal{L}$ refers to a low-pass filtering operation. However, we do not have access to the isolated contribution of the star, $\textbf{d}_s$, at the location of the companion, but only the total signal, $\textbf{d}$. Rewriting Eq. \ref{eq:base_model}, while ignoring the systematics, and plugging this into Eq. \ref{eq:alpha_est} we have:
\begin{equation}
    \hat{\alpha}(\lambda) =\frac{\mathcal{L}  \left[ \textbf{d}- \textbf{d}_p\right]}{\mathcal{L} \left[ \textbf{f}_s \right] }.
\end{equation}
For a linear low-pass filter we can split these terms up as:
\begin{equation}\label{eq:alpha_est2}
 \hat{\alpha}(\lambda)  = \frac{\mathcal{L}[\textbf{d}]}{\mathcal{L}[\textbf{f}_s]} - \frac{\mathcal{L}[\textbf{d}_p]}{\mathcal{L}[\textbf{f}_s]} 
 .\end{equation}
In other words, we have to correct for the planet continuum leaking into the estimate of $\alpha$, which effectively functions as a highpass filter on the planet model. This correction on the planet signal was included in the forward model of the planet signal, which is discussed in the next subsection and ignored here. We used a second-order Savitsky-Golay filter with a kernel width of 301 pixels as our low-pass filter.

Additionally, because CRIRES is a slit spectrograph, the line spread function (LSF) can change along the slit. For example, in good seeing conditions with adaptive optics, the full width at half maximum (FWHM) of the point spread function (PSF) is smaller than the width of the slit. This means that on-axis the spectral resolution of the stellar contribution is determined by the width of the PSF, while off-axis this is determined by the width of the slit. We account for this effect by including shifted versions of the stellar model in the linear model matrix, which allows for flexibility in fitting the LSF of the stellar contribution.

Considering both effects, we included the following components in $M_{\psi}$ for the stellar contribution to the observed data:

\begin{equation}\label{eq:lsf}
    \mathbf{m}_{s, k} =  \frac{\mathcal{L}[\textbf{d}]}{\mathcal{L}[\textbf{f}_s]} \textbf{f}_s[\lambda + k\Delta \lambda],
\end{equation}
where $\textbf{f}_s[\lambda + k\Delta \lambda]$ refers to the stellar master spectrum shifted by $k$ pixels to account for the changing LSF along the slit. We used integer shifts for $k$ from -3 to 3 wavelength bins throughout this work. 
%\begin{equation}
%    \mathbf{d}_{s, k} =  \frac{\mathcal{L}[D(x, \lambda)]}{\mathcal{L}[M_s(\lambda)]} M_s(\lambda + k\Delta \lambda)
%\end{equation}

\subsubsection{Planet contribution}
The observed planet signal can be written as:
\begin{equation}
    \textbf{d}_p = c_p T(\lambda) F_p(\lambda, \psi),
\end{equation}
where $T(\lambda)$ is the transmission of both the atmosphere and the instrument and includes the unit conversion from flux density to detector counts; $F_p(\lambda, \psi)$ is a model of the planetary spectrum with nonlinear parameters, $\psi$; and $c_p$ is a unitless linear scaling parameter that is fitted for. The planetary emission model and its parameterization are described in Sect. \ref{sec:atmosphere_models}. We estimated the transmission using the stellar master spectrum and a PHOENIX model of $\beta$ Pic ($F_s(\lambda)$):
\begin{equation}
    T(\lambda) = \frac{\textbf{f}_s}{F_s(\lambda)}.
\end{equation}
Considering that the continuum of the planet model leaks into the estimate of $\alpha$ (Eq. \ref{eq:alpha_est2}) and, subsequently, the stellar model, we have to correct for this in our model of the planet contribution. Effectively, this applies a high-pass filter to the planet contribution and leads to following corrected planetary contribution, $\textbf{d}_p'$:
\begin{equation}
    \begin{split}
   \textbf{d}_p' &= \textbf{d}_p -  \frac{\mathcal{L}[\textbf{d}_p]}{\mathcal{L}[\textbf{f}_s]}\textbf{f}_s \\
    &= c_p \left( T(\lambda) F_p(\lambda, \psi) -  \frac{\mathcal{L}[T(\lambda) F_p(\lambda, \psi)]}{\mathcal{L}[\textbf{f}_s]}\textbf{f}_s \right).
    \end{split}
\end{equation}

We therefore included the following linear component for the corrected planet signal in our model matrix $M_{\psi}$:
\begin{equation}
  \mathbf{m}_p(\psi) =T(\lambda) F_p(\lambda, \psi) -  \frac{\mathcal{L}[T(\lambda) F_p(\lambda, \psi)]}{\mathcal{L}[\textbf{f}_s]}\textbf{f}_s.
\end{equation}
We note that we also have to correct for the planet continuum leaking into the shifted versions of the stellar spectrum used for modeling the LSF (Eq. \ref{eq:lsf}). However, since the amplitude of the central stellar component dominates over the amplitudes corresponding to the shifted versions, we ignore this effect throughout this work.

\subsubsection{Systematics and noise}
After an initial subtraction of the stellar contribution from each row, we found from our visual inspection that there  still was some correlated structure in the data. This may, for example, originate from the imperfect removal of the stellar and telluric contamination or other systematics. To remove most of these correlated residuals, we included a model of the systematics in our forward model:
\begin{equation}
    \textbf{d}_{\textrm{sys}} = \sum_i^N c_i \textbf{m}_{\textrm{sys},i},
\end{equation}
where $c_i$ are the linear coefficients and $\textbf{m}_{\textrm{sys},i}$ the systematics modes. This systematics model was computed the same way as in \citet{2020AJ_wilcomb_kappa_and_b_osiris} and \citet{Ruffio_2021AJ_HR8799_deep}. First, we subtracted the estimated stellar contribution from each row by fitting the model without the planet contribution. Subsequently, we scaled the residuals of each row to the same local continuum by dividing by the low-pass filtered data. The systematics were then determined from these residuals using a principal component analysis (PCA), while excluding the four spaxels around the considered position of the companion. This was done to avoid the planet signal leaking into the systematics model. Finally, the PCA components were scaled back to the local continuum at the considered location. We used ten PCA components in the systematics model throughout this work. To remove any residual outliers, we did an initial subtraction of the star and systematics model and masked 4$\sigma$ outliers. This leaves us with mainly uncorrelated noise, which we checked by inspecting the autocorrelation function of the residuals. 

We also found the noise estimates from the pipeline to not be sufficiently accurate, which is crucial for obtaining realistic uncertainties on retrieved atmospheric parameters. Instead, we estimated the noise as follows: For each exposure, the stellar and systematics models were constructed and subtracted from the observed data. The noise was then estimated by taking the standard deviation over time of these residuals. These uncertainties are used as diagonal entries for the covariance matrix $\Sigma_0$. Following \citet{2019AJ_ruffio_radial_velocity}, we then include a scaling parameter, $s$, to account for a possible underestimation of these uncertainties: $\Sigma = s^2 \Sigma_0$, which is fitted for using the data.

\subsection{Likelihood}\label{sec:likelihood}

Following \citet{2019AJ_ruffio_radial_velocity}, we can write our forward model in matrix form:
\begin{equation}
    \textbf{d} = M_{\psi}\textbf{c} + \eta
,\end{equation}
where the model matrix $M_{\psi}$ is given by:
\begin{equation}
M_{\psi}= \left[
        \mathbf{m}_p(\psi), 
        \mathbf{m}_{s,-3\Delta \lambda},  \dots,
        \mathbf{m}_{s,3\Delta \lambda},
        \mathbf{m}_{\textrm{sys}, 0}, \dots,
        \mathbf{m}_{\textrm{sys}, 10}  \right],
\end{equation}
$\textbf{c}$ are the  linear coefficients and $\eta$ is again the uncorrelated noise. Following the derivation from  \citet{2019AJ_ruffio_radial_velocity}, which uses a prior of $P(s) \propto s^{-\gamma}$ for the noise scaling parameter, $s$, we can analytically marginalize over the linear parameters $\mathbf{c}$ and noise scaling $s$. This results in the following posterior:
\begin{equation}\label{eq:likelihood}
    L(\psi|\textbf{d}) = \frac{P(\psi)}{\sqrt{|\Sigma_0| \times |M_\psi^T \Sigma_0^{-1} M_\psi|}} \left( \frac{1}{\chi_0^2} \right)^{\frac{N_d - N_c + \gamma - 1}{2}},
\end{equation}
with $P(\psi)$ as the prior on the planetary parameters, $\psi$; $N_d$ is the number of data points; $N_c$ is the number of components in the linear model, and
\begin{equation}
    \chi_0^2 = (\textbf{d} - M_{\psi} \hat{\textbf{c}})^T \Sigma_0^{-1}(\textbf{d} - M_{\psi} \hat{\textbf{c}}),
\end{equation}
where $\hat{\textbf{c}}$ is the linear least squares solution such that:
\begin{equation}
    \hat{\textbf{c}}^T M_{\psi}^T \Sigma_0^{-1}M_{\psi} = \textbf{d}^T \Sigma_0^{-1}M_{\psi},
\end{equation}
which we find using a numerical least squares solver. For details on the derivation we refer the reader to Appendix D from \citet{2019AJ_ruffio_radial_velocity}. We used $\gamma=2$ and calculated the likelihood for each nodding position and night separately and added the log-likelihoods. We did this to avoid additional interpolation of the data, as the wavelength solution is slightly different for the different nodding positions and nights. Furthermore, we applied Eq. \ref{eq:likelihood} to each spectral order separately to avoid flux calibration errors between spectral orders having a major impact on the result. This means we are effectively fitting for the continuum of the planet. While this discards information on the continuum, we found that this can lead to more robust results with respect to particular choices in the data reduction. This also means that the noise scaling factor is allowed to deviate per spectral order.

\section{Atmospheric models}\label{sec:atmosphere_models}
\subsection{Nominal model}
We used \textit{petitRADTRANS} \citep{2019A&A_molliere_prt, 2020A&A_molliere_hr8799} to generate model spectra of $\beta$ Pictoris b. We followed a similar modeling approach as in, among others, \citet{Gravity2020A&A_betapicb} and \citet{2021Natur_zhang_isotope, 2021A&A_zhang_isotope_bd}. Our nominal model uses a free pressure-temperature (P-T) profile, equilibrium chemistry, and the cloud model from \citet{2020A&A_molliere_hr8799}. The P-T profile consists of four free knots located at 1 mbar, 0.1 bar, 1 bar, and 10 bar, respectively. The full P-T profile was then calculated using third-order B-spline interpolation in log-space of pressures and temperature. A prior on the temperature was chosen such that it smoothly decreases towards lower pressures and no inversions are possible. The chemical equilibrium model is detailed in \citet{Molliere2017A&A_jwst_chemistry}, which takes as input the P-T profile, metallicity [Fe/H], C/O ratio, and quenching pressure, $P_{\textrm{quench}}$, and returns the abundances of all the species. We include Rayleigh scattering from H$_2$, He, collisional induced absorption of H$_2$-H$_2$ and H$_2$-He and line-by-line opacities of H$_2$O \citep{Polyansky2018MNRAS_waterlinelist}, CO \citep{Rothman2010JQSRT_hitemp} and CH$_4$ \citep{Hargreaves2020ApJS_methanelinelist}. The cloud model is presented in \citet{2020A&A_molliere_hr8799} and is based on the EddySed model from \citet{AckermanMarley2001ApJ_clouds}. We included opacity sources of Fe and MgSiO$_3$ clouds, assuming crystalline, irregularly-shaped particles, and account for scattering in the radiative transfer. We chose these cloud species as they are likely the most prominent at the expected temperature of $\beta$ Pictoris b and were retrieved in its atmosphere by \citet{Gravity2020A&A_betapicb}. Additionally, we fitted for the radial velocity (RV) of the planet and the line broadening caused by the spin of the planet. For the rotation of the planet we used the \textit{fastRotBroad} function from pyAstronomy \citep{Czesla2019ascl.soft_pyastronomy}, which has as its free parameters: the projected spin velocity, $v\sin(i)$, and linear limb darkening coefficient, $\epsilon$.

Initially, we found the posteriors of the surface gravity, metallicity and P-T profile to be strongly correlated and unstable to changes in the modeling or data reduction, which influenced the constraints on the other parameters. This issue of trying to constrain the surface gravity from just K-band spectroscopy has been noted in other works \citep{2021Natur_zhang_isotope, PalmaBifani2023A&A_ABPic_sinfoni}. To relieve this issue we use a Gaussian prior on the surface gravity based on the measurement of the dynamical mass from \citet{Lacour2021A&A_betapicb_gravity} of 11.9 $\pm$ 3.0 $M_{\textrm{Jup}}$ and chose a radius prior of 1.4 $\pm$ 0.1 $R_{\textrm{Jup}}$ based on the values of \citet{Chilcote2017AJ_beta_pic_b_gpi, Stolker2020A&A_miracles, Gravity2020A&A_betapicb}, which gives $\log(g) = 4.18 \pm 0.13$. The priors on the parameters are summarized in Table \ref{tab:priors}. 
\begin{table}[]
\center
\renewcommand{\arraystretch}{1.15}
\caption{Parameters used in the atmospheric models of the planet with their priors.}
\label{tab:priors}
\begin{tabular}{l|l|l}
\hline

Parameter                       & Prior        & Models     \\ \hline
{[}Fe/H{]} or {[}C/H{]}                       & U(-1.5, 1.5)  & a,b,c,d    \\
C/O                             & U(0, 1)       & a,c,d  \\
$\log(g)$ {[}cgs{]}                & G(4.18, 0.13) & a,b,c,d    \\
$T_0$ {[}K{]} (10 bar)          & U(1500, 4000) & a,b,c   \\
$T_1$ {[}K{]} (1 bar)           & U(0.5, 1) * T$_0$ & a,b,c\\
$T_2$ {[}K{]} (0.1 bar)         & U(0.5, 1) * T$_1$ & a,b,c\\
$T_3$ {[}K{]} (1 mbar)          & U(0.5, 1) * T$_2$ & a,b,c\\
RV {[}km/s{]}                   & U(20, 40)  & a,b,c,d          \\
$v\sin(i)$ {[}km/s{]}            & U(0, 40)  & a,b,c,d     \\
Limb darkening $\epsilon$              & U(0, 1)  & a,b,d         \\
$\log (P_{\textrm{quench}})$ & U(-3, 2) & a,b,c,d \\
$\log(X_0^{\textrm{MgSiO}_3})$ & U(-3, 1)  & a,b,d     \\
$\log(X_0^{\textrm{Fe}})$        & U(-3, 1)     & a,b,d     \\
$f_{\textrm{sed}}$              & U(0, 10)     & a,b,c,d     \\
$\log(K_{zz})$                    & U(5, 13)      & a,b,c,d    \\
$\sigma_g$                      & U(1.05, 3)     & a,b,c,d  \\
$\log (\tau_{\textrm{cloud}})$  & U(0, 2)    & c \\
$ \log (P_{\textrm{cloud, base}})$ & U(-1, 0.5) & c
\end{tabular}
    \begin{tablenotes}
          \small
          \item \textbf{Notes.} The last column shows the models that include the specified parameter. Models: a) Nominal model, b) Free composition, c) Forced grey cloud deck, d) \citet{Gravity2020A&A_betapicb} prior. For the last model (d), the priors consist of Gaussians with uncertainties specified in \citet{Gravity2020A&A_betapicb} and not the priors listed here, except for the high-resolution parameters (RV, $v\sin(i)$, $\epsilon$).
    \end{tablenotes}
\end{table}

\begin{figure*}[htbp]
    \centering
    \includegraphics[width=\linewidth]{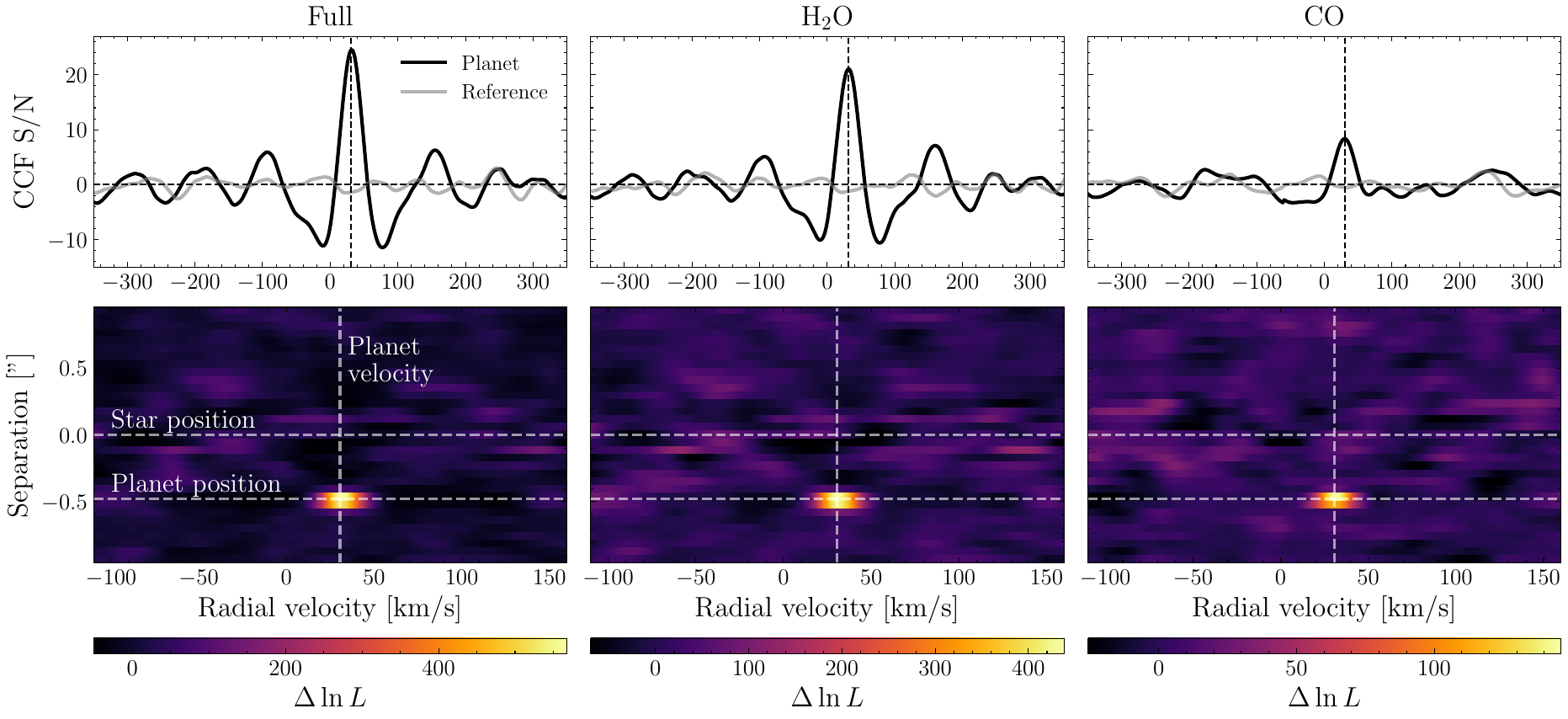}
    \caption{Detection of the planet using model templates of the full planet, H$_2$O and CO using cross-correlation (top row) and the likelihood ratio (bottom row). The radial velocity is shown here in the barycentric restframe. In the top panel we also show the cross-correlation value for a reference spectrum, which was taken at the same separation on the other side of the star. In the bottom panels we indicate the location of the star and the planet and the expected velocity of the planet.}
    \label{fig:detection_maps}
\end{figure*}

\subsection{Alternative models}
To test the robustness of the results to model assumptions, we also ran retrievals with alternative models. We tested the following alternative models:

    \textbf{Free composition}: Same as the nominal model, but now with free composition, where we fit for the logarithm of the abundance of the species, which is assumed to be constant vertically in the atmosphere. We also include minor species (CO$_2$, HCN, NH$_3$ and $^{13}$CO) in these retrievals. This is done to search for the presence of these species in a more robust way than the crude cross-correlation analysis  (detailed in Section \ref{sec:detection}) and allow us to obtain upper limits on the abundance of these species. The retrieved abundances are subsequently converted to a C/O ratio and metallicity.
     
     \textbf{Forced grey cloud deck}: In Section \ref{sec:results}, we show that the nominal model lead to cloud-free solutions, which is inconsistent with the low spectral resolution data at shorter wavelengths \citep{Chilcote2017AJ_beta_pic_b_gpi,Gravity2020A&A_betapicb}. To test whether this has impact on the values of the other retrieved parameters we include a strong prior on the cloudiness of the planet. Instead of determining the cloud location and opacity from the condensation curves, we follow a similar approach as in \citet{2017MNRAS_burningham_retrieval_Ldwarfs} and fit for the location and opacity of the cloud deck, for which we chose priors such that the cloud is optically thick and in the photosphere of the planet. This alleviates the uncertainty in the composition of the clouds, which can impact the location of the cloud deck.

     \textbf{Strong prior based on low-resolution data}: We use the posteriors obtained by \citet{Gravity2020A&A_betapicb} as the prior here and fix the P-T profile to the one obtained in \citet{Gravity2020A&A_betapicb}. This is similar to jointly fitting the GPI, GRAVITY, and our new CRIRES+ data with equal weighting for the combined low-resolution data and the high-resolution data.

\section{Results}\label{sec:results}
\subsection{Detection of molecules}\label{sec:detection}
First, we aim to detect molecules in the atmosphere of the planet. We do this by calculating the log-likelihood (as defined in Eq. \ref{eq:likelihood}) as a function of position along the slit and radial velocity. The templates used are generated with the nominal model discussed in Section \ref{sec:atmosphere_models} with the parameters found in \cite{Gravity2020A&A_betapicb} and $v\sin(i)=25$ km/s \citep{2014Natur_snellen_beta_pic_crires}. In the case of the full model, we compare the likelihood ratio with respect to a model not including any planet signal, so only the stellar model and systematics. For the molecules, we evaluated the likelihood ratio between the full model and the likelihood without the considered species. For example, for water, we have:
\begin{equation}
    \Delta \ln L_{\textrm{H$_2$O}} = \ln L_{\textrm{full}} -  \ln L_{\textrm{without H$_2$O}}.
\end{equation}

The results are shown in Fig. \ref{fig:detection_maps}, showing clear detections of H$_2$O and CO. The planet signal is predominantly detected in two spaxels at a roughly equal strength, which we simply summed for the remainder of the analysis. We also show the cross-correlation function (CCF) for the combined planet signal following \citet{2019AJ_ruffio_radial_velocity}, given by the linear parameter corresponding to the planet component in the model while including only the specified species in the planet model. We normalized the CCF by the standard deviation of the CCF of the combined two spaxels at the same separation as the planet but on the other side of the star. Using this, we estimate detection signal-to-noise ratios (S/Ns)  of 24.5, 21.0, and 8.4 for the full model, water, and carbon monoxide, respectively. We also searched for features of CO$_2$, CH$_4$, HCN, and NH$_3$ and we did not find any significant signal, as expected from equilibrium chemistry and the S/N of our observations. The high S/N detections are a major improvement over the results from \citet{2014Natur_snellen_beta_pic_crires}, who were unable to detect water using the original CRIRES. In fact, while not shown here, we are able to detect the planet confidently ($\mathrm{S/N} > 5$) in almost all individual exposures of two minutes. This improvement in data quality is mainly due to significantly increased instantaneous wavelength coverage of the instrument and improved detectors. A visual comparison between our cross-correlation map and the one from \citet{2014Natur_snellen_beta_pic_crires} is shown in Fig. \ref{fig:crires+comp}.

%\begin{figure*}
%    \centering
%    \includegraphics[width=\linewidth]{beta_pic_comparison.pdf}
%    \caption{}
%    \label{fig:crires+comp}
%\end{figure*}

\subsection{Atmospheric retrievals}
\begin{figure}
    \centering
    \includegraphics[width=\linewidth]{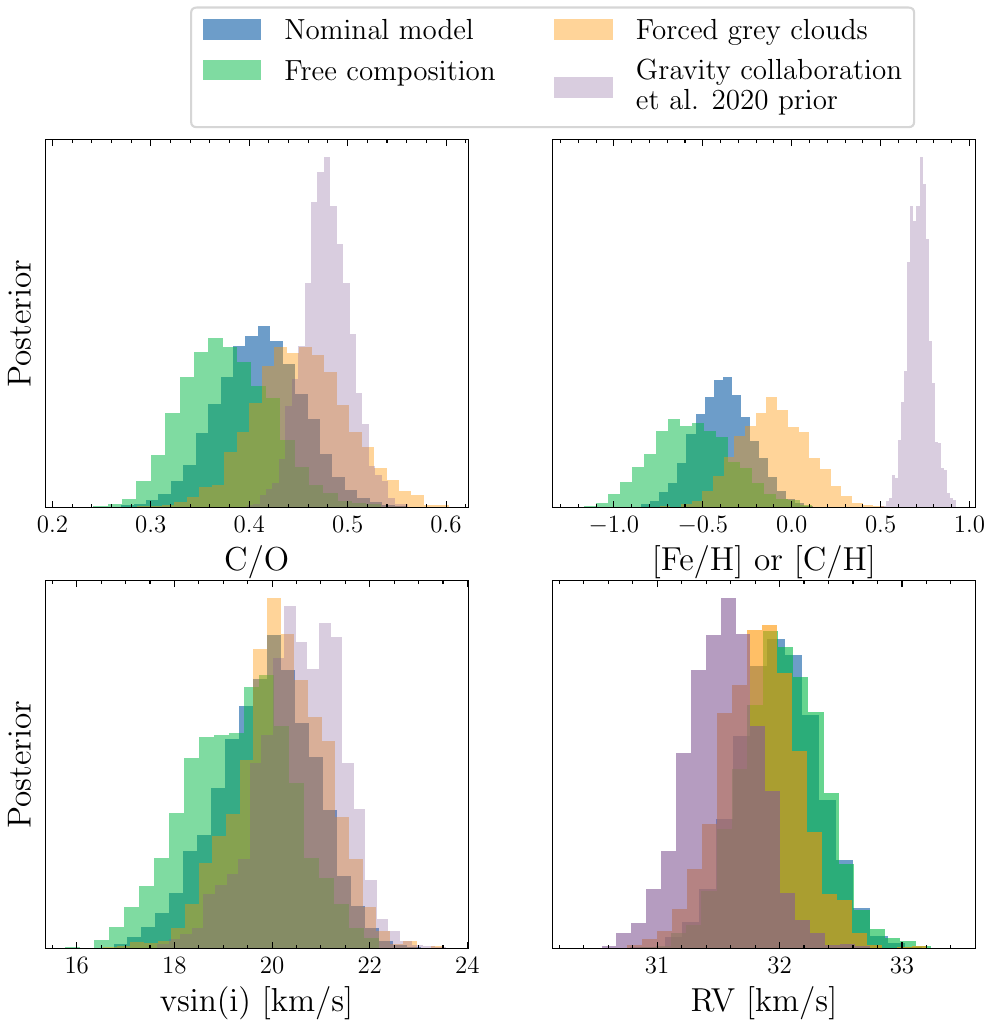}
    \caption{Constraints on a subset of the parameters from the atmospheric retrievals using the models discussed in Section \ref{sec:atmosphere_models}. The RV shown here is in the barycentric restframe.}
    \label{fig:retrieved_params}
\end{figure}

\begin{figure}
    \centering
    \includegraphics[width=\linewidth]{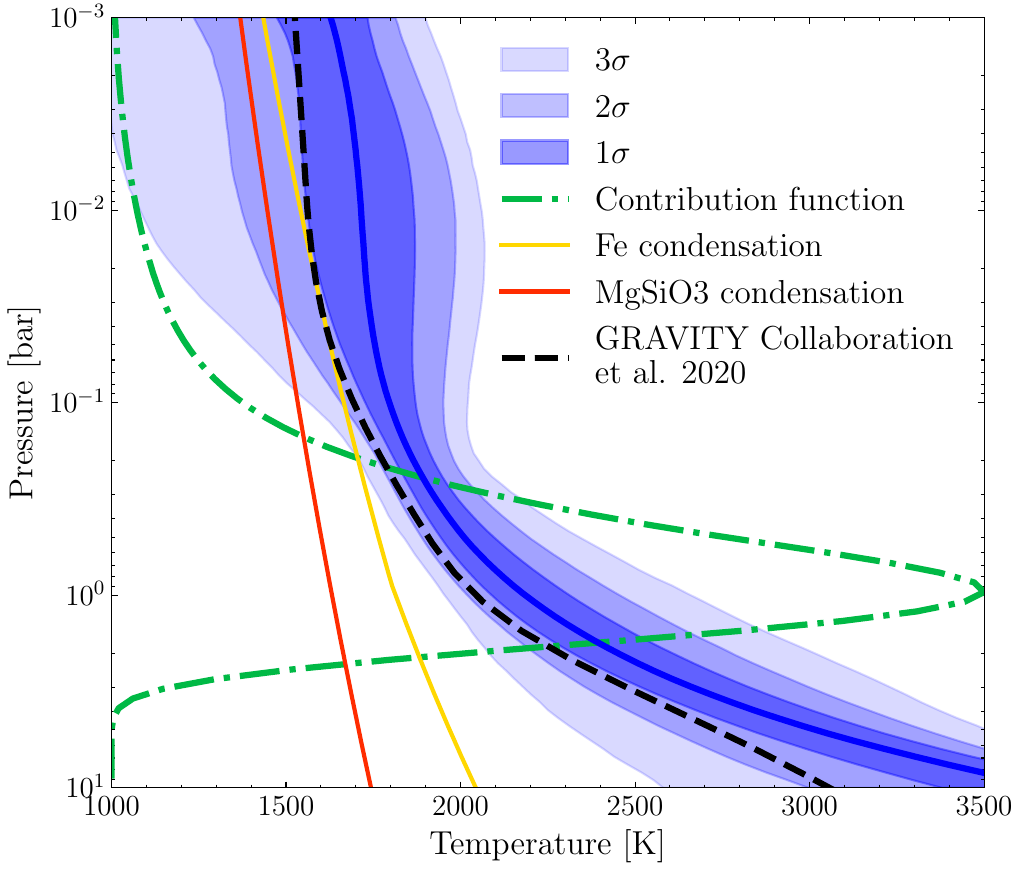}
    \caption{Retrieved P-T profile for the nominal model with associated uncertainties. Also shown: the P-T profile obtained by \citet{Gravity2020A&A_betapicb}, the condensation curves of Fe and MgSiO$_3$ for the best fitting parameters, and the emission contribution function for the best fitting parameters.}
    \label{fig:retrieved_PT}
\end{figure}

We ran atmospheric retrievals on the CRIRES+ data for the models specified in Section \ref{sec:atmosphere_models}. We sampled the posterior using \textit{PyMultiNest} \citep{Buchner2016ascl.soft_pymultinest}, a Python wrapper for MultiNest \citep{Feroz2019OJAp_MultiNest}, with 500 live points and an initial sampling efficiency of 80\%, which is the recommended value for parameter estimation. The obtained posterior distribution for a selection of parameters is listed in Table \ref{tab:posteriors} and shown in Fig. \ref{fig:retrieved_params}. The full corner plot for the nominal model is shown in Fig. \ref{fig:full_corner}. The values of the retrieved parameters are discussed in the next section. The retrieved P-T profile for the nominal model is shown in Fig. \ref{fig:retrieved_PT}, together with the condensation curves for the best fitting parameters. 

\begin{table*}[htbp]
\centering
\caption{\centering Retrieved posterior values for the main parameters of interest.}
\label{tab:posteriors}
\renewcommand{\arraystretch}{1.2}
\begin{tabular}{l|cccc}

\hline
Model               & C/O          & [Fe/H] or [C/H] &  RV [km/s]  & vsin(i) [km/s] \\ \hline
Nominal                    & $0.41 \pm 0.04$  & $-0.39 \pm 0.16$ & $31.99 \pm 0.32$& $20.2 \pm 1.0$  \\
Free composition & $0.38 \pm 0.05$& $-0.49 \pm 0.25$ & $31.90 \pm 0.33$  & $19.9 \pm 1.0$  \\
Forced clouds  & $0.45 \pm 0.05$ & $-0.10 \pm 0.20$ & $31.85 \pm 0.32$  & $ 19.9 \pm 1.1$  \\
Low-resolution prior & $0.48 \pm 0.03$  & $0.72 \pm 0.06$ &$31.55 \pm 0.30$ & $20.6 \pm 0.8$       \\ \hline    
\end{tabular}
\end{table*}

\begin{figure*}
    \centering
    \includegraphics[width=0.9\linewidth]{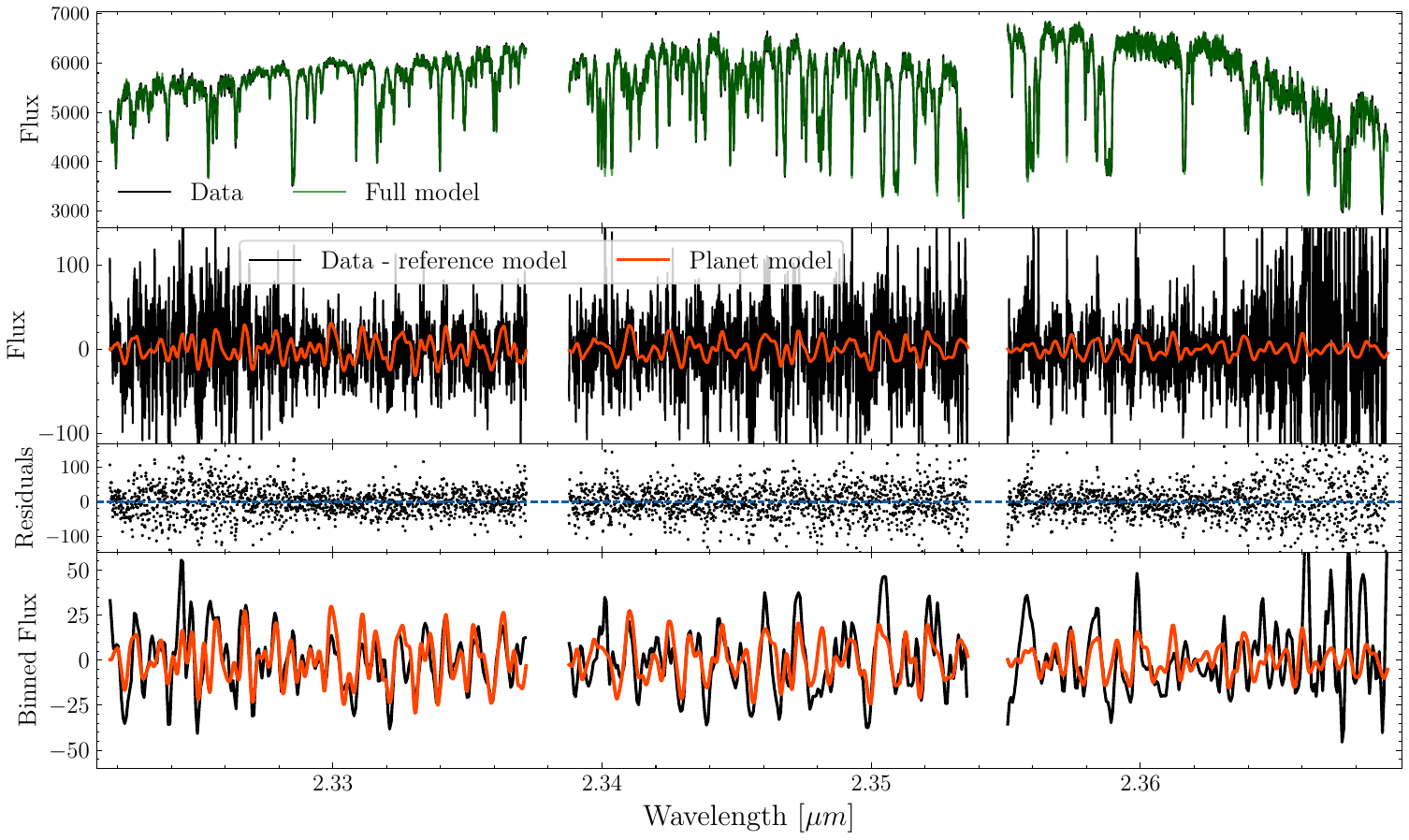}
    \caption{CRIRES+ data from one of the spectral orders along with the fitted model. Top panel shows the total extracted flux at the location of the planet with the full fitted model including star, planet, and systematics. The second panel shows the data after subtracting the stellar and systematic components, along with the best-fitting planet model. The third panel shows the residuals of this fit. Finally, the bottom panel shows the same as the second panel but then smoothed to a spectral resolution of 15,000 to show the broad planet features more clearly.}
    \label{fig:data_fit}
\end{figure*}

To test whether our models provide a good fit to the data, we show the spectral order with the highest detection significance together with the fitted model in Fig. \ref{fig:data_fit}. While the data are  noisy at the native resolution, the bottom panel in Fig. \ref{fig:data_fit} shows the data smoothed to a spectral resolution of 15,000, which is the intrinsic resolution of the planet signal set by its rotational velocity. This shows good agreement between the data and our model, allowing us to clearly identify individual absorption lines. Finally, we compare the spectral energy distribution of the retrieved models with the low spectral resolution Y-, J-, and H-band data from GPI \citep{Chilcote2017AJ_beta_pic_b_gpi} and K-band data from \citet{Gravity2020A&A_betapicb} in Fig. \ref{fig:comp_lowres}. The data from the different bands are scaled according to the retrieved scaling parameters from \citet{Gravity2020A&A_betapicb}. The models are generated using the correlated-k mode of petitRADTRANS and include additional opacities from TiO, FeH, VO, \textbf{Na} and K. Since we did not fit for the radius, the models were scaled to match the K-band flux of the GPI data. The figure shows that all models provide a good fit to the data in the K-band, but that there is a large deviation at shorter wavelengths.

\section{Discussion}\label{sec:discussion}
\subsection{Abundance ratios and metallicity}
For the nominal model, we obtained a sub-solar C/O ratio of $0.41\pm 0.04$, which is consistent with the values obtained by \citet{Gravity2020A&A_betapicb}. The alternative models result in similar constraints, albeit slightly higher for cloudy models. For the free composition model, the measured C/O ratio is the ratio in the gas phase only. In the presence of condensation, a part of the oxygen will be sequestered into clouds and the true C/O ratio will be even lower. While the uncertainty on the C/O ratio is likely underestimated, the retrieved value was found to be relatively robust to changes in the data reduction parameters and modeling choices. This illustrates the power of high-resolution spectroscopy for determining robust elemental abundance ratio's of exoplanets. The interpretation of this sub-solar C/O ratio is limited by the fact that the oxygen abundance in the host star is unknown. It is therefore hard to make any statements on the formation and migration history of the planet. In the case the C/O ratio of $\beta$ Pictoris is solar, \citet{Gravity2020A&A_betapicb} showed the sub-solar value for $\beta$ Pictoris b, in combination with its high mass, points towards a formation through core accretion and not gravitational collapse. 

We obtained a metallicity of $-0.39 \pm 0.16$ for the nominal model, which is significantly lower than the value of $0.66 \pm 0.13$ found by \citet{Gravity2020A&A_betapicb} on the combined GPI Y, J , and H data with the GRAVITY K-band data, but similar to the value of $-0.53 \pm 0.3$ they obtained using only the GRAVITY K-band data. The derived metallicity is also strongly dependent on the used model, shown by the large deviations in Fig. \ref{fig:retrieved_params}. This is the result of a degeneracy between clouds and metallicity, which  is discussed in the next subsection. We note that constraining absolute abundances and thus the planets metallicity using high-resolution spectroscopy is relatively hard, especially when one can not retain the planets continuum. Combining high-spectral-resolution data with low resolution over a larger wavelength range may help relieve these issues. In our case, this is shown from the retrieval with the priors from \citet{Gravity2020A&A_betapicb}, where the metallicity converges to its prior due to the limited information present in the high-resolution spectra.

For the free composition retrieval, we only obtain upper bounds on the minor species. There is a slight preference for a model with $^{13}$CO, but we consider this marginal and do not discuss it further in this paper.

\subsection{Clouds and P-T profile}
\begin{figure}[htbp]
    \centering
    \includegraphics[width=\linewidth]{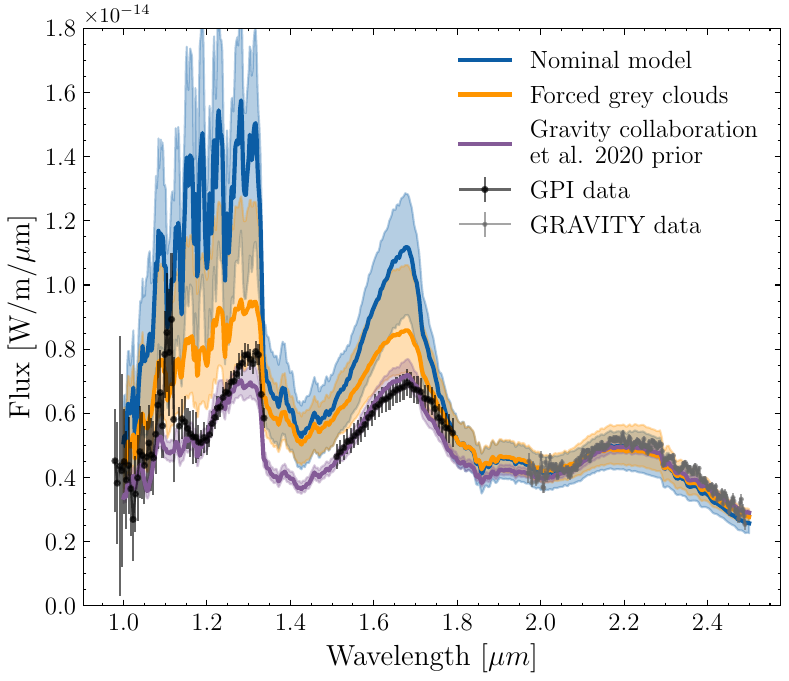}
    \caption{Comparison of the spectral energy distributions of our retrieved models compared to the low-resolution GPI data from \citet{Chilcote2017AJ_beta_pic_b_gpi} and GRAVITY data from \citet{Gravity2020A&A_betapicb}. Since we do not retrieve radius, the flux of our models is scaled such that it matches the GPI flux in the K-band. The solid line shows the median value and the colored region shows the 16th and 84th percentiles.}
    \label{fig:comp_lowres}
\end{figure}

The retrieval for the nominal model results in no constraints on the cloud parameters. This is the result of the retrieved P-T profile not intersecting the condensation curves of the considered cloud species for the retrieved parameters. The retrieved P-T profile is in shape similar to the one obtained by \citet{Gravity2020A&A_betapicb} but shifted to slightly higher temperatures. This is because high-resolution spectroscopy, after high-pass filtering, is mostly sensitive to the slope of the P-T profile and not its absolute value. Furthermore, the derived metallicity is much lower, which results in lower condensation temperatures. These effects together lead to cloud-free solution that are inconsistent with the low-resolution data at shorter wavelengths, as shown in Fig. \ref{fig:comp_lowres}. 

\begin{figure*}[htbp]
    \centering
    \includegraphics[width=0.9\linewidth]{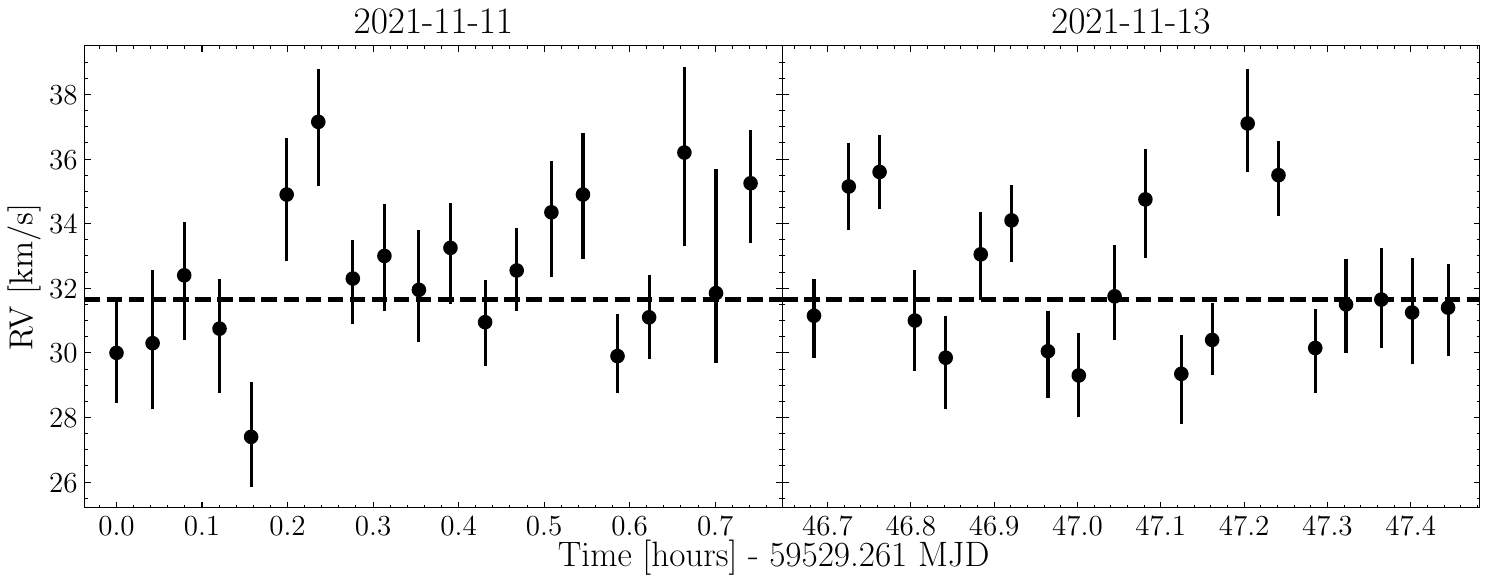}
    \caption{Estimation of the radial velocity in the barycentric restframe for individual exposures.}
    \label{fig:rv_over_time}
\end{figure*}

High-resolution K-band retrieval converging to cloud-free solutions has been seen in other works \citep[e.g.,][]{2021A&A_zhang_isotope_bd, Xuan2022ApJ_browndwarf_kpic} and may be the result of the wavelength coverage in combination with the parameterization and priors on the P-T profile. We have tried the P-T parameterization of \citet{2020A&A_molliere_hr8799} and obtained almost equivalent results to the nominal model. The alternative model with a forced grey cloud deck already agrees better with the low-resolution data. Naturally, the retrieval with the results from \citet{Gravity2020A&A_betapicb} as a prior agrees even better with the low-resolution data. Still, the main parameters of interest for this work, C/O, $v\sin(i)$ and RV, seem to be minimally affected by the presence of the clouds, as shown in Fig. \ref{fig:retrieved_params}. The derived metallicity is highly correlated with the presence of clouds and changes depending on the modeling choices. This may be caused by the degeneracies between the fitted coefficient of the planet component, clouds, and metallicity, as all these parameters mainly change the depth of the spectral lines. Our observations also do not help constrain the composition and location of the clouds in the atmosphere of $\beta$ Pic b. Observations over a larger wavelength range with e.g. JWST could resolve this uncertainty, as shown by the recently detected silicate cloud absorption features in VHS 1256-1257 b \citep{Miles2023ApJ_vhs_jwst}.

\subsection{Radial velocity and spin rotation}
We obtained a radial velocity of $31.9 \pm 0.3$ km/s in the barycentric rest frame. Using the systematic velocity, $v_\mathrm{sys} = 20 \pm 0.7$ km/s \citep{Gontcharov2006AstL...32..759G_rv_stras} for $\beta$ Pictoris, this gives a radial velocity between the star and the planet of $11.9 \pm 0.8$ km/s, where the uncertainty is dominated by the uncertainty on the radial velocity of the star. This is consistent at $\approx$2$\sigma$ with the radial velocity, $v_\mathrm{orb} = 10.0 \pm 0.1$ km/s, predicted from the orbital solution \citep{Wang2021ascl.soft_whereistheplanet}. Since we are able to detect the planet in individual frames, we can study the RV evolution over time (shown in Fig. \ref{fig:rv_over_time}), exhibiting no clear variability. We find a rotational velocity $v\sin(i)= 19.9 \pm 1.0$ km/s. This is slightly lower than the value obtained in \citet{2014Natur_snellen_beta_pic_crires}, but consistent within 2$\sigma$. Using a radius of $1.4 \pm 0.1$ $R_{\textrm{Jup}}$ and inclination of $88.95 \pm 0.10$ degrees, this gives a rotation period of $8.7 \pm 0.8$ hours for $\beta$~Pic~b, assuming that the planet has no obliquity. We note that the limb darkening coefficient is converging to its maximum value in the retrievals. Since there is a correlation between the limb darkening and $v\sin(i)$, the rotational velocity could be lower if the limb darkening coefficient turns out to be lower.

\section{Summary and outlook}\label{sec:conclusions}
We report on observations of $\beta$ Pictoris b using the recently upgraded CRIRES+. We present dedicated features in \textit{pycrires} for the reduction of observations of directly imaged planets with this instrument. By forward-modeling the data in combination with free atmospheric retrievals, we attempted to characterize the atmosphere of the planet. We find a slightly sub-solar C/O ratio, $v\sin(i) = 19.9 \pm 1.0$ km/s, which gives a rotation period of $8.7 \pm 0.8$ hours and a radial velocity of $31.9\pm 0.3$ km/s in the barycentric restframe. However, the results from the nominal retrieval on solely the CRIRES+ data are inconsistent with the low-resolution data at shorter wavelengths. By either forcing the presence of clouds or using the results from \citet{Gravity2020A&A_betapicb} as a prior we obtain solutions that are more consistent with the low-resolution data. Fortunately, the main parameters of interest are minimally affected by this. This shows the power of high-resolution spectroscopy in breaking degeneracies in atmospheric retrievals and deriving robust elemental abundance ratios. Furthermore, we are able to confidently detect the planet in individual exposures of two minutes, showing the significantly improved capabilities of CRIRES+. Searching for exomoons using radial velocity or monitoring the spectral variability of some of the most favourable objects may thus already be within reach, especially with the arrival of HiRISE \citep{Vigan2023arXiv230912390V_hirise}, which couples the high-contrast imager SPHERE to CRIRES+.

\begin{acknowledgements}
R.L., I.S., S.d.R. acknowledge funding from the European Research Council (ERC) under the European Union's Horizon 2020 research and innovation program under grant agreement No 694513. T.S.\ acknowledges the support from the Netherlands Organisation for Scientific Research (NWO) through grant VI.Veni.202.230. A.S.L. acknowledges financial support from the Severo Ochoa grant CEX2021-001131-S funded by MCIN/AEI/ 10.13039/501100011033. This work used the Dutch national e-infrastructure with the support of the SURF Cooperative using grant no. EINF-4556.

\end{acknowledgements}

\bibliographystyle{aa}
\bibliography{references}

\appendix

\section{Visual comparison with Snellen et al. 2014}\label{app:crires_comp}
\begin{figure*}[htbp]
    \centering
    \includegraphics[width=\linewidth]{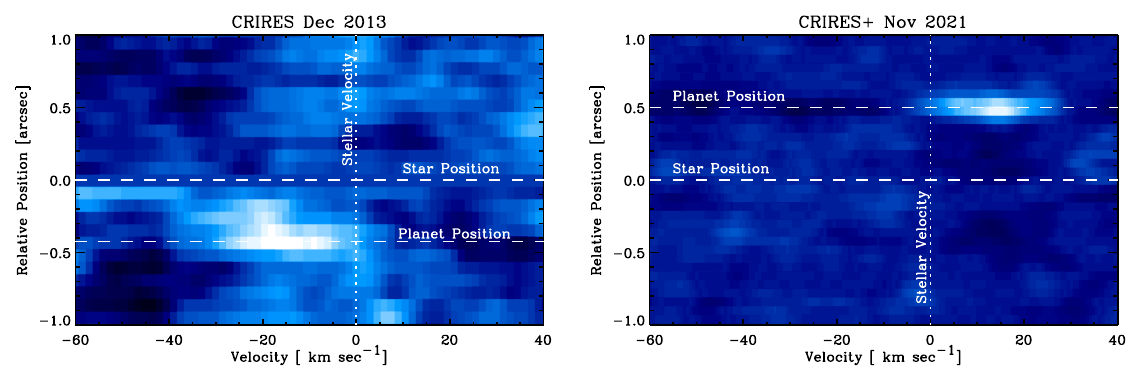}
    \caption{Visual comparison between the cross-correlation signal-to-noise ratio map of $\beta$ Pic b using the original CRIRES \citep{2014Natur_snellen_beta_pic_crires} and our new CRIRES+ observations. The relative position along the slit was matched to the convention used in \citet{2014Natur_snellen_beta_pic_crires}.}
    \label{fig:crires+comp}
\end{figure*}

\section{Full corner plot for the nominal model}\label{app:corner_plot}
\begin{figure*}[htbp]
    \centering
    \includegraphics[width=\linewidth]{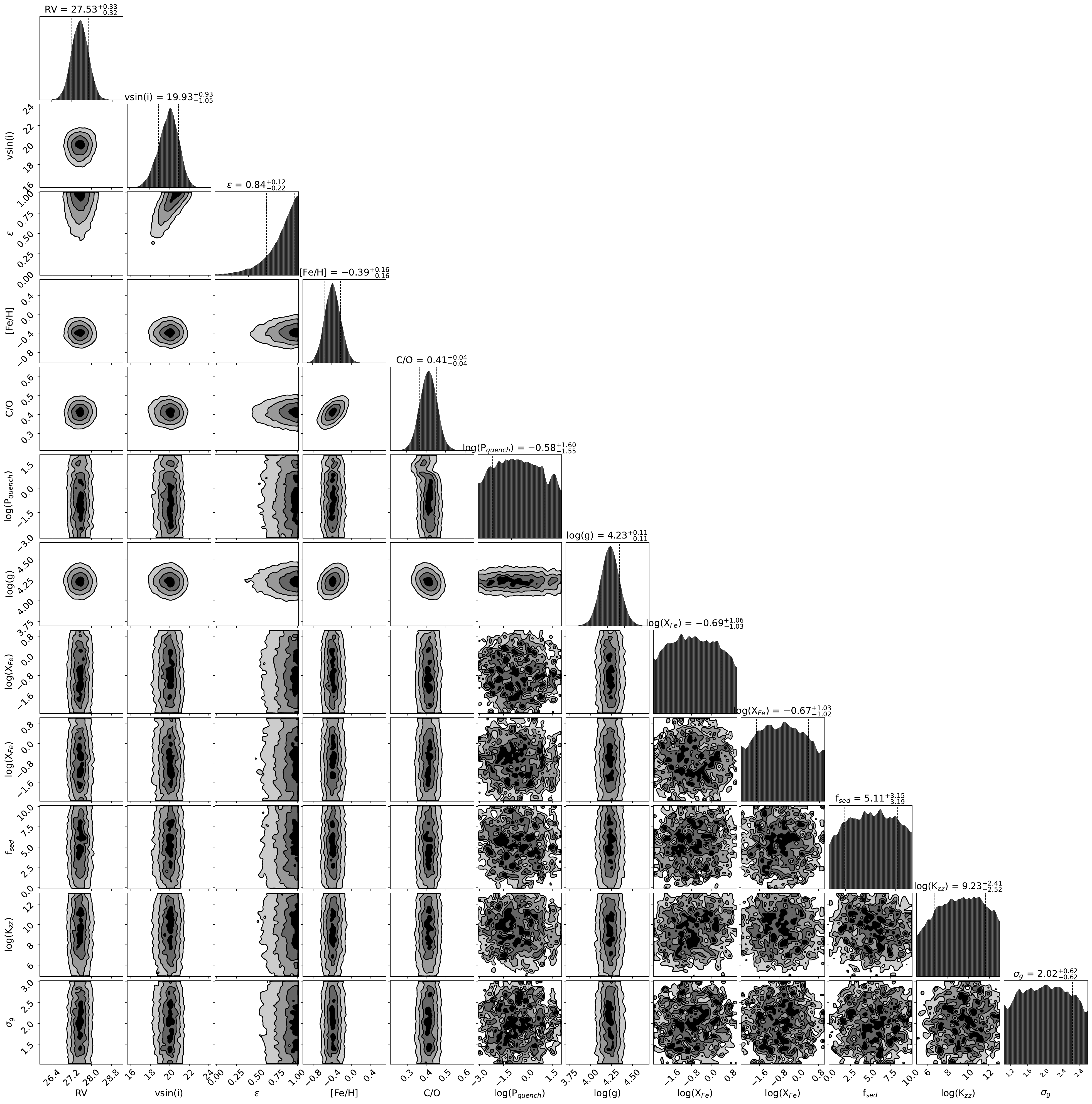}
    \caption{Full corner plot of the atmospheric retrieval for the nominal model.}
    \label{fig:full_corner}
\end{figure*}

\end{document}